\documentclass[12pt]{article}
\usepackage{a4wide,amsmath,amssymb,alltt,graphicx,enumitem}
\usepackage{colordvi}
\graphicspath{{screenshots/}}

\newcommand\ie{i.e.\ }
\newcommand\eg{e.g.\ }
\newcommand\rd{\mathrm{d}}

\newcommand\button[1]{\lower .45ex\hbox{\includegraphics[height=2.2ex]{button_#1}}}
\newcommand\TM{\texttrademark}
\newcommand\ri{\mathrm{i}}

\makeatletter
\def\reportno#1{\gdef\@reportno{#1}}
\def\@maketitle{%
  \hfill{\small\begin{tabular}[t]{r}%
    \@reportno
  \end{tabular}\par}%
  \vskip 2em%
  \begin{center}%
    \let \footnote \thanks
    {\large \@title \par}%
    \vskip 1.5em%
    {%\large
      \lineskip .5em%
      \begin{tabular}[t]{c}%
        \@author  
      \end{tabular}\par}%
    \vskip 1em%
    {%\large
     \@date}%
  \end{center}%
  \par
  \vskip 1.5em}
\makeatother

\begin{document}

\reportno{CERN-TH--2017--175, CP3-17-26, MaPhy-AvH/2017--07, \\
MSUHEP--17--013, MPP--2017--173}

\title{Loopedia, a Database for Loop Integrals}

\author{%
  C.~Bogner$^1$, S.~Borowka$^2$, T.~Hahn$^3$, G.~Heinrich$^3$,
  S.P.~Jones$^3$, M.~Kerner$^3$, \\
  A.~von~Manteuffel$^4$, M.~Michel$^5$,
  E.~Panzer$^6$, V.~Papara$^3$ \\[1ex]
${}^1$ Institut f\"ur Physik, Humboldt-Universit\"at zu Berlin,
  D--10099 Berlin, Germany \\[1ex]
${}^2$ Theoretical Physics Department, CERN, Geneva, Switzerland \\[1ex]
${}^3$ Max-Planck-Institut f\"ur Physik \\
  F\"ohringer Ring 6, D--80805 Munich \\[1ex]
${}^4$ Department of Physics and Astronomy \\
  Michigan State University, East Lansing, MI 48824, USA \\[1ex]
${}^5$ CP3, Universit\'e Catholique de Louvain,
  B--1348 Louvain-la-Neuve, Belgium \\[1ex]
${}^6$ All Souls College, University of Oxford,
  OX1 4AL, Oxford, UK}

\date{\today}

\maketitle

\begin{abstract}
\noindent
Loopedia is a new database at \texttt{loopedia.org} for information on 
Feynman integrals, intended to provide both bibliographic information as 
well as results made available by the community.  Its bibliometry is 
complementary to that of \textsc{spires} or arXiv in the sense that it 
admits searching for integrals by graph-theoretical objects, \eg its
topology.
\end{abstract}

%========================================================================

\section{Introduction}

Researchers in high-energy physics enjoy nearly unsurpassed (in both 
quality and freedom of access) bibliographic access and resources 
through well-maintained databases such as \textsc{spires} and arXiv.  
For the practitioner in loop calculations one recurring problem is, 
however, that the existing databases are indexed solely by `traditional' 
metrics: author, title, year of publication, etc.  It is not even 
possible to formulate queries of the kind ``Find publications which 
refer to loop integral $X$,'' where $X$ is specified in some 
graph-theoretical way, say by its topology.

Loopedia attempts to fill this gap, though it is not limited to 
bibliographic information.  The description field of each record can 
hold any kind of textual information (\eg URLs to software), and in 
addition arbitrary files can be uploaded, for example Fortran programs 
or Maple worksheets.

What we presently provide and describe herein is the database software 
only, with a few entries for illustration.  Populating the database with 
content is a job both beyond the capabilities of a trained librarian and 
ultimately in the interest of the researchers who computed the loop 
integrals, to make their contributions easily findable and get cited.

Integrals are generally associated with their underlying graph.  In the 
case of non-trivial numerators, linear combinations, or set integrals 
for a sector, the reference is indexed by the corner (scalar) integral 
of the sector (topology).  Tensor integrals (in the one-loop case) are 
intentionally not covered; justifiable exceptions can be put with the 
scalar integral, with a suitable explanation.

%========================================================================

\section{Prolegomena}

For the purposes of Loopedia an $L$-loop Feynman integral is usually an
object of the form
\begin{equation}
\int\frac{\rd^D\ell_1\cdots\rd^D\ell_L}
  {(k_1^2 - m_1^2)^{\nu_1}\cdots(k_n^2 - m_n^2)^{\nu_n}},
\end{equation}
where the $k_i$ are linear combinations of the loop momenta $\ell_1, 
\dots, \ell_L$ and the graph's external momenta $q_1, \dots, q_E$, 
dimensionally regularized in $D$ dimensions.  (Exceptions to this form 
can be accomodated as long as the integral possesses a corresponding
Feynman graph.)

Feynman loop integrals are rendered as graphs and as such can be 
identified by the Nickel index \cite{nagel,nickel,graphstate}.  The 
latter is the central object by which integrals are indexed in Loopedia 
(user input may be in the form of an adjacency list, too).  The Nickel 
index is arguably the most compact representation of a graph and 
constructed as follows: start at a vertex $v$ and write down all 
vertices connected to $v$ (\texttt{e} for external legs), then insert a 
`\texttt{|}'.  Repeat for the other vertices but omit edges already 
included.

Besides this `bare' Nickel index, which represents just the graph's 
topology, Loopedia introduces an augmented form, the colored Nickel 
index, \textsl{CNickel} for short, which additionally captures the 
layout of masses and external $q^2$.  Loopedia color-codes them on 
screen:

\begin{gather*}
\begin{aligned}
\text{Nickel~~} &
\lower .5ex\hbox{\includegraphics[height=2.2ex]{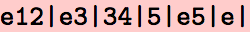}} \\
\text{CNickel~~} &
\lower .5ex\hbox{\includegraphics[height=2.2ex]{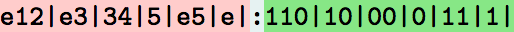}}
\end{aligned}
\qquad
\lower 6ex\hbox{\smash{\includegraphics[height=15ex]{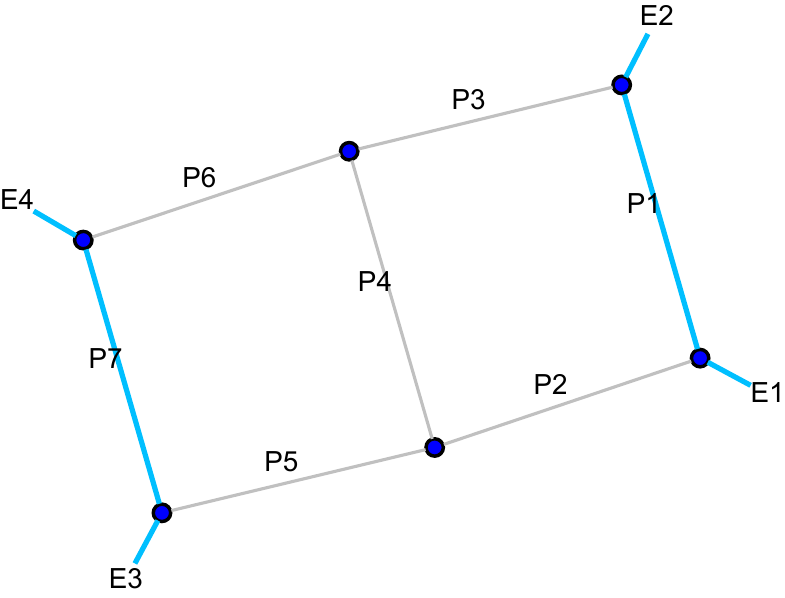}}}
\end{gather*}
For each `edge identifier' of the Nickel (\texttt{e} or a digit), the 
coloring has a mass identifier:
\begin{itemize}[leftmargin=8em, itemsep=0pt, parsep=0pt, topsep=0pt]
\item[\texttt{0}~]
	for zero,
\item[\{\texttt{z}, \texttt{n}, \texttt{s}\}~]
	for \{any, non-zero, special\} mass scale,
\item[\texttt{1}\dots\texttt{9}, \texttt{a}\dots\texttt{y}~]
	(not \texttt{n}, \texttt{s})
	for a definite non-zero mass scale.
\end{itemize}
The `\texttt{s}' option is for cases like thresholds or 
pseudo-thresholds, and the exact meaning of `special' should be noted in 
the description of the integral.

The `definite' identifiers also represent arbitrary scales (though not zero) 
but unlike the `\texttt{zns}' choices they express equality of scales within 
the graph, like named patterns in a computer-algebra system.  The graph in 
the example above has several massive propagators/legs, but all with the same 
scale `\texttt{1}'.

%========================================================================

\begin{figure}[ht]
\centerline{\includegraphics[width=.75\hsize]{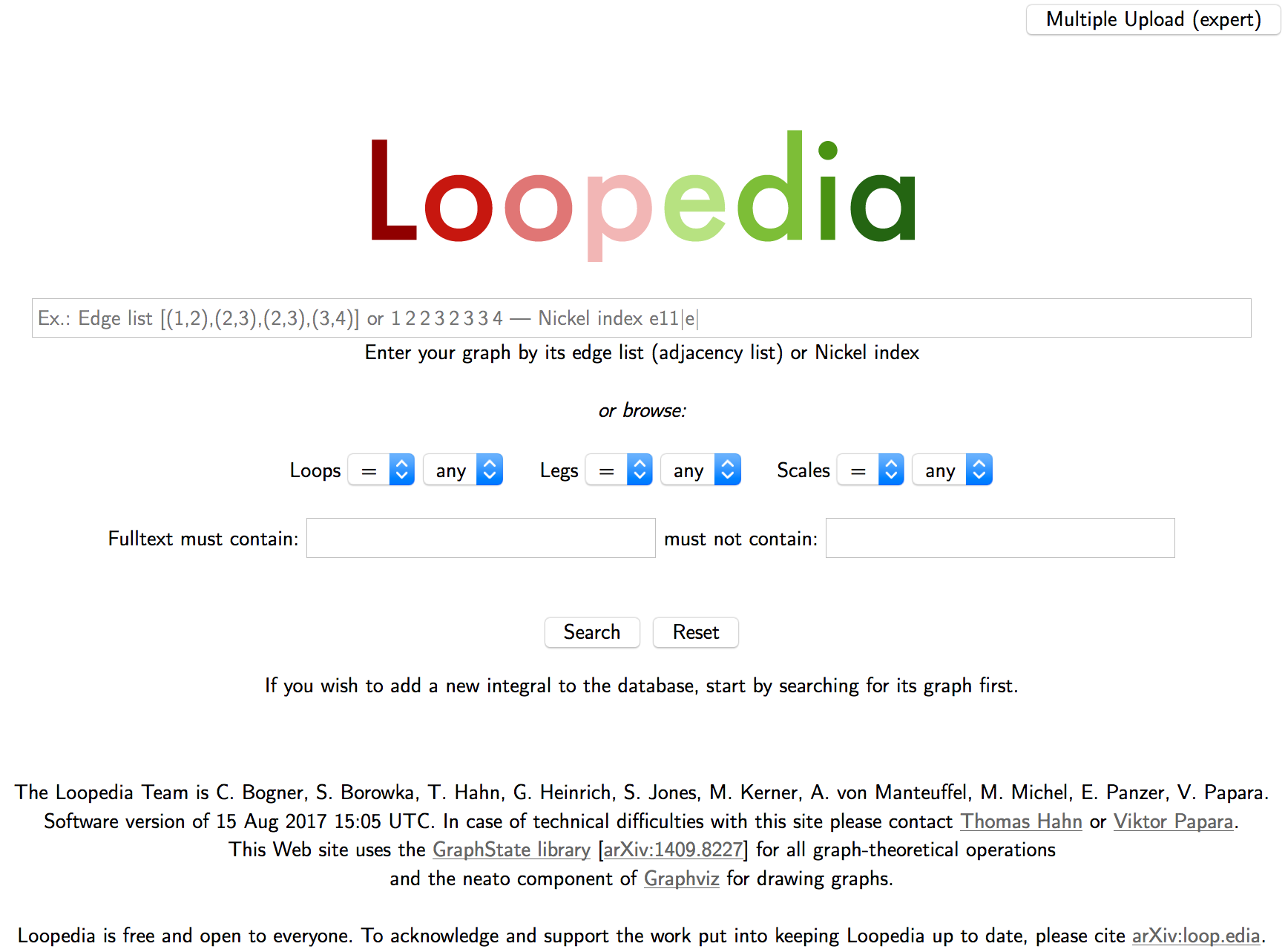}}
\caption{\label{fig:startpage}The Loopedia Start Page}
\end{figure}

\section{User Guide}

Loopedia is located at
$$
\texttt{loopedia.org}
$$
and upon access, the Start Page\TM\ (Fig.~\ref{fig:startpage}) is 
displayed. One enters a graph in the upper input field and/or chooses 
constraints for the search with the lower controls and then hits the 
\button{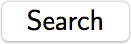} button.

While the Nickel index is a very compact way of denoting a graph, it may 
not be the notation most people are used to and also few graph-theory 
software packages are able to output Nickel indices.  Loopedia therefore 
admits a graph's \textsl{edge list (adjacency list)} as input, too.  The 
typist has some latitude there: generally the graph is understood as 
long as a pairing of vertices into propagators can reasonably be made 
out.  This means that \eg FeynArts \cite{feynarts} or QGRAF \cite{qgraf} 
notation can directly be pasted into the input form.  An edge list is 
converted to a Nickel index right away.

All [C]Nickel indices are first brought into a standard form (for 
details see Sect.~\ref{sect:graphth}).  Because the standard form 
involves amputation of a tadpole's single leg, searching for one leg 
automatically redirects to zero legs.

\begin{figure}[ht]
\centerline{\includegraphics[width=.7\hsize]{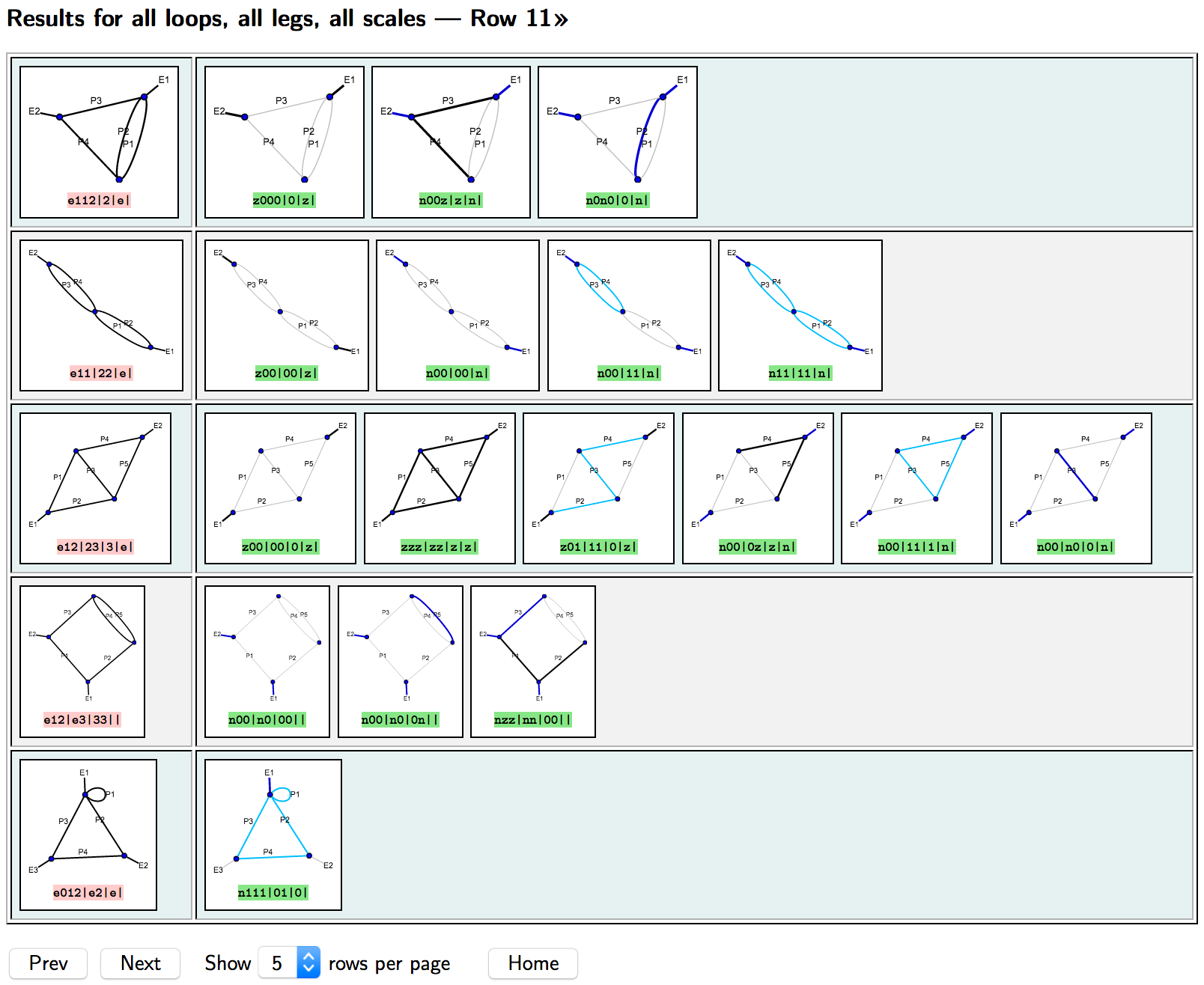}}
\caption{\label{fig:browse}The Loopedia Graph Browser}
\end{figure}

%========================================================================

\subsection{Browsing Graphs}

The search results are displayed in Loopedia's Graph Browser\TM\ 
(Fig.~\ref{fig:browse}).  The matches are organized so that the topology 
appears on the left side and the configurations for which information is 
available are on the right.  Every graph icon is clickable.

\begin{figure}[ht]
\centerline{\includegraphics[width=.7\hsize]{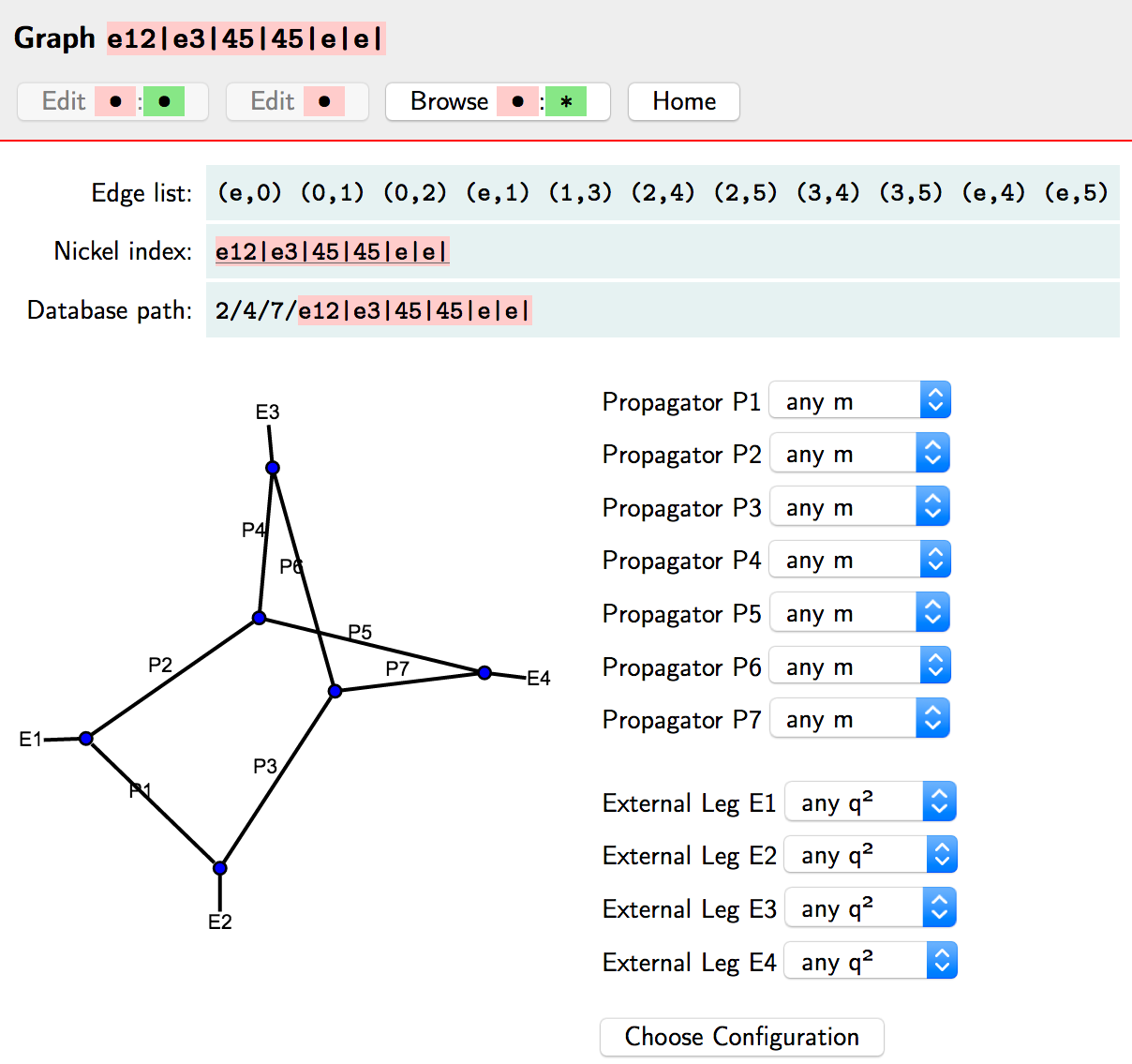}}
\caption{\label{fig:confedit}The Loopedia Configuration Editor}
\end{figure}

\begin{figure}[ht]
\centerline{\includegraphics[width=.85\hsize]{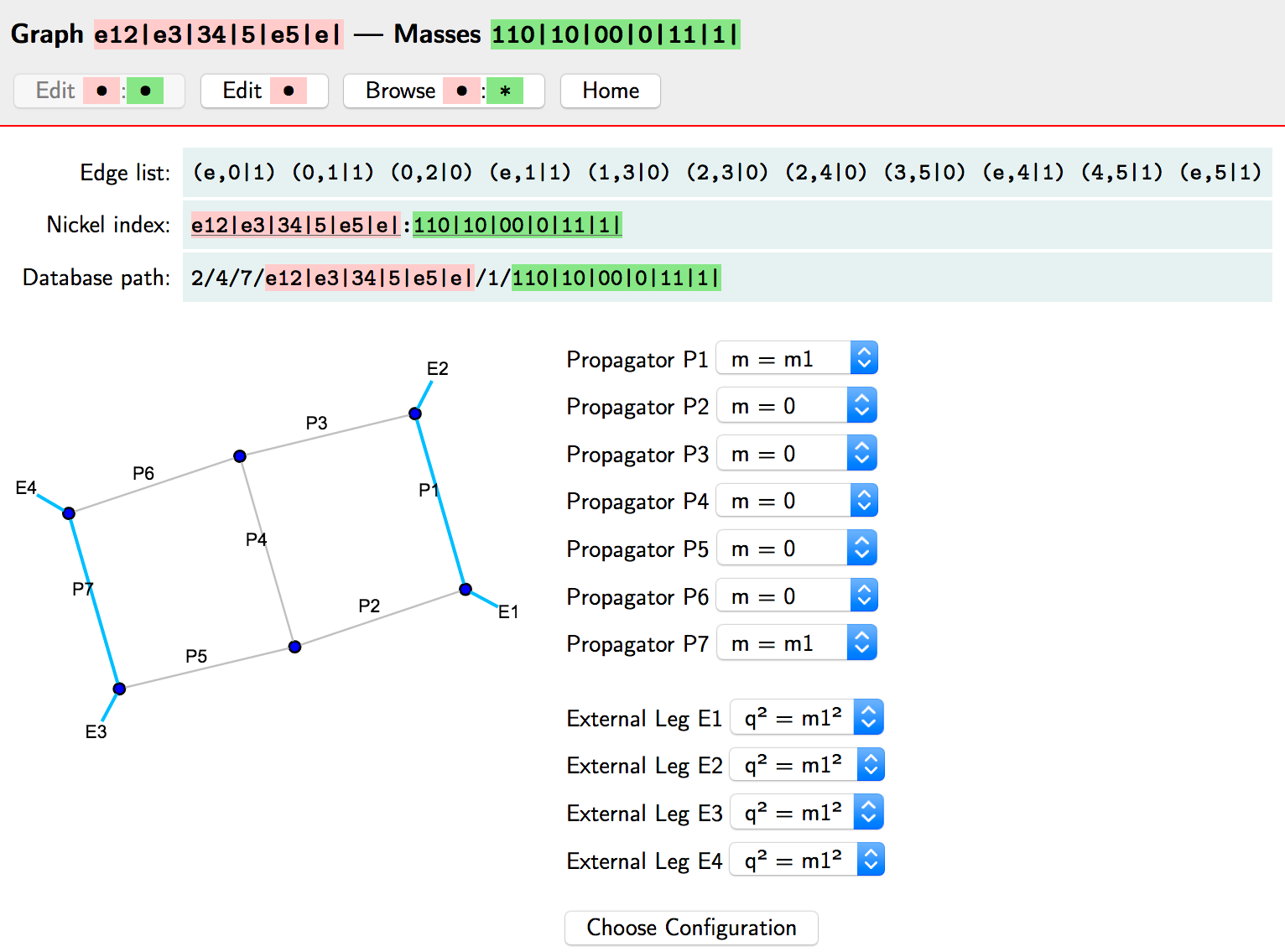}}
\caption{\label{fig:recdisp}The Loopedia Record Display}
\end{figure}

Clicking on one of the graph icons opens the single-graph display for 
that graph.  For icons in the left column it starts the Configuration 
Editor\TM\ (Fig.~\ref{fig:confedit}), for icons in the right column it 
opens the Record Display\TM\ (Fig.~\ref{fig:recdisp}).  The Graph 
Browser is skipped and the Configuration Editor started immediately if 
the search turns up no matches and an explicit graph was entered, to 
commence adding a new record.

%========================================================================

\subsection{Viewing Single Graphs}

The navigation panel at the top has buttons for editing the 
configuration \button{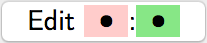}, the topology \button{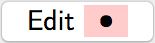}, 
browsing the topology's configurations \button{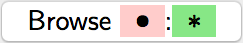}, and 
returning to the start page \button{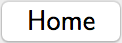}.  Depending on the situation 
not all buttons may be active.

In the main part of the page, a picture of the graph is shown underneath 
its various textual representations (edge list, [C]Nickel index, 
database path).  To the right is the Configuration Chooser\TM, with the 
defaults preset to the current values (if any).

\begin{figure}[ht]
\centerline{\includegraphics[width=.3\hsize]{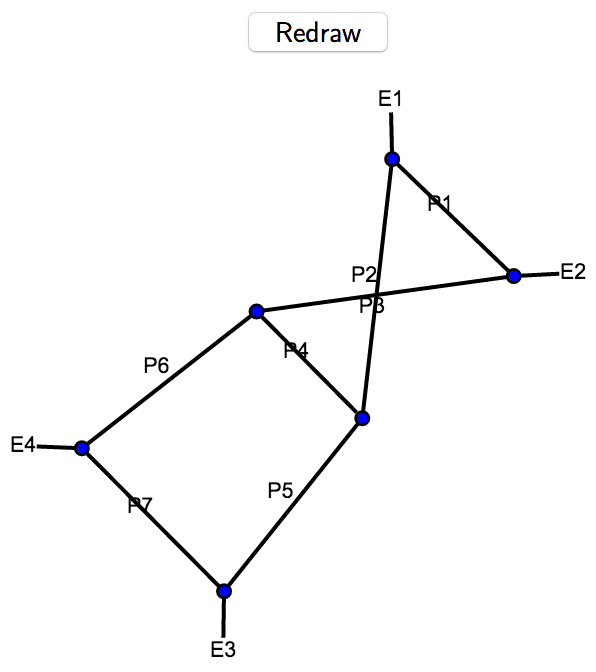}}
\caption{\label{fig:redraw}An `ugly' graph and the Redraw button.}
\end{figure}

From time to time a graph will have an `ugly' rendering.  If it is the 
first diagram of its kind (\ie no records added yet) the Configuration 
Editor will feature a `Redraw' button on top of the graphic 
(Fig.~\ref{fig:redraw}) to draw it again with a different random seed.  
When entering new graphs, please allow your aesthetic subconcious to 
press that button if necessary!

\begin{figure}[ht]
\centerline{%
\parbox[t]{.1985\hsize}{\hrule height 0pt width 0pt
  \includegraphics[width=\hsize]{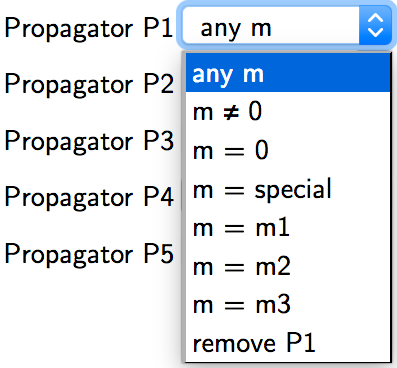}}\qquad\quad
\parbox[t]{.209\hsize}{\hrule height 0pt width 0pt
  \includegraphics[width=\hsize]{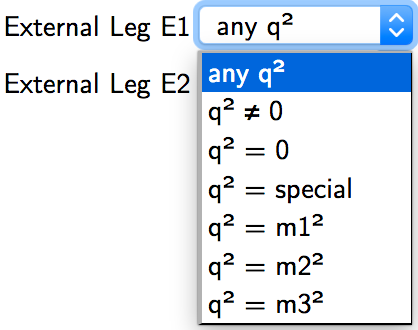}}}%
\caption{\label{fig:confmenus}Drop-down Menus for Graph Configuration}
\end{figure}

The menu items refer to the labelling of the graph: P1, P2, etc.\ for 
internal propagators and E1, E2, etc.\ for external legs.  
Fig.~\ref{fig:confmenus} shows the drop-down menus expanded out.  
Observe that the `\textsf{remove P$n$}' option in the mass menu is 
somewhat of an anomaly as it removes the corresponding propagator, \ie 
it changes not the colored but the bare Nickel index.

Each $m_i$ listed in the menus represents an arbitrary non-zero scale.  
Unlike the `any' choice they allow to express equality of scales within 
the graph, like named patterns in a computer-algebra system.  The actual 
names are insignificant and will in general be renumbered in the colored 
Nickel index constructed from the HTML form.  All different $m_i$ is the 
most general case; choosing identical $m_i$ or zero refers to a special 
case.  For a loop integral with three propagators, for example, any of 
the following combinations identifies the case of the first two masses 
equal: $(m_1,m_1,m_2)$, $(m_2,m_2,m_1)$, $(m_2,m_2,m_3)$, 
$(m_3,m_3,m_2)$, $(m_1,m_1,m_3)$, $(m_3,m_3,m_1)$.

\begin{figure}[ht]
\centerline{\includegraphics[width=.9\hsize]{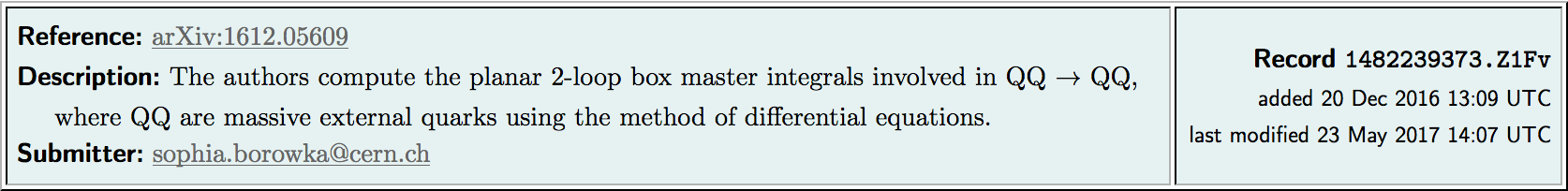}}
\caption{\label{fig:record}Example of a single record listing}
\end{figure}

\begin{figure}[ht]
\centerline{\includegraphics[width=.8\hsize]{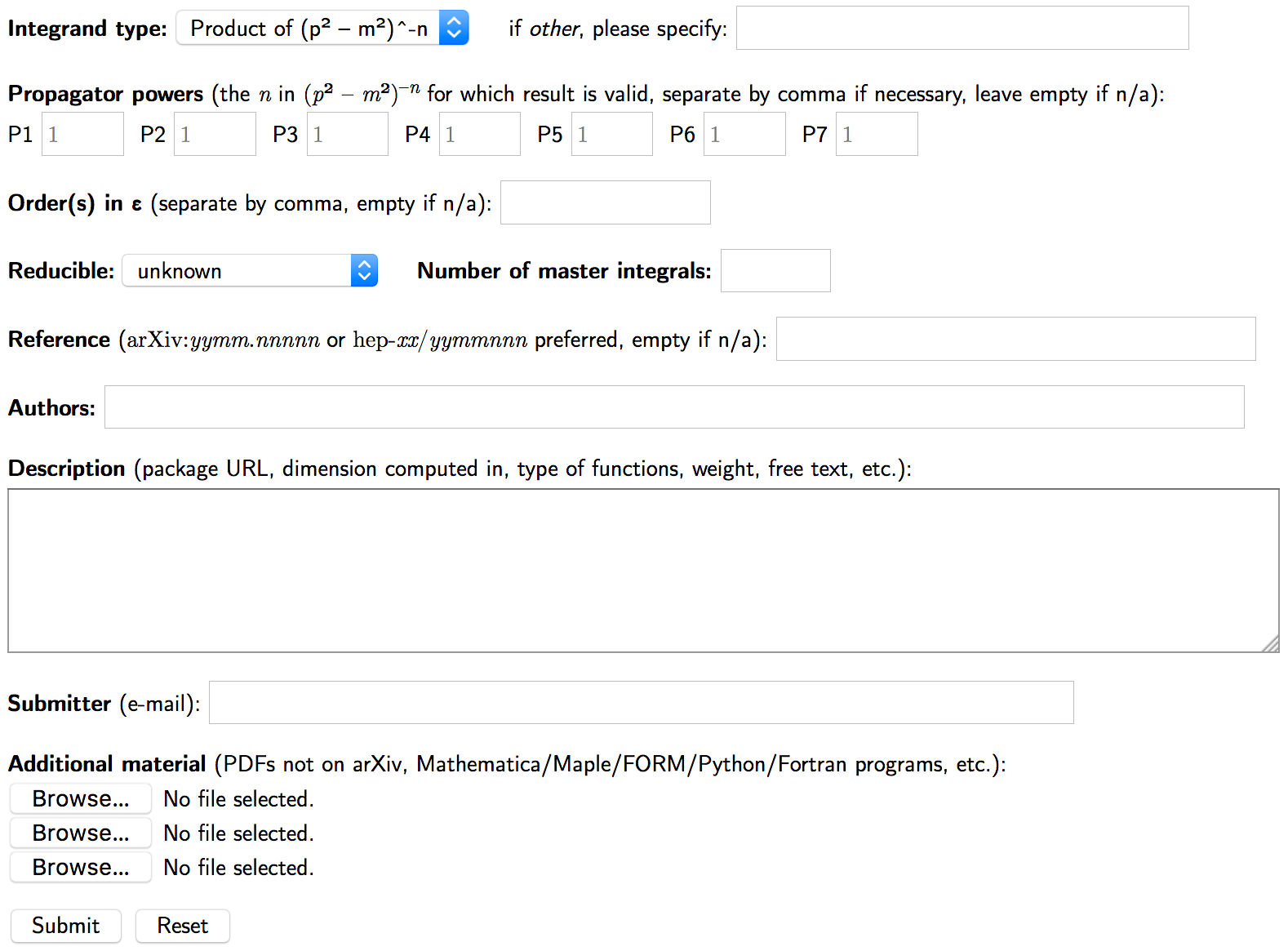}}
\caption{\label{fig:newrec}The New Record Form}
\end{figure}

When editing a topology (bare Nickel index only) the page ends here, \ie
in order to proceed one needs to choose a configuration first.  
Otherwise the page continues with the available records 
(Fig.~\ref{fig:record}) and the New Record Form\TM\ 
(Fig.~\ref{fig:newrec}).

%========================================================================

\subsection{Entering New Records}

To enter a new record one must first navigate to a graph and 
configuration as outlined above.  The New Record Form is found beneath 
the available records.  Its fields are largely optional although at 
least either the Reference or Description field must be non-empty.

While a few of the input fields ask for specific information on the 
integral we have found ourselves unable to come up with a (reasonably 
concise) list of fields that would cover all cases, given the diversity 
of strategies and concepts followed in the computation of loop 
integrals.  For this reason a special importance falls onto the 
Description field as general-purpose free-text field.  Things we 
encourage submitters to put in the description field include
\begin{itemize}
\item Detailed bibliographic information.  State in which equation or 
section the integral is given.  Provide hints about auxiliary
information, \eg definitions relevant to the integral.  Point to 
ancillary files or resources, \eg included in the arXiv submission.
In general: try to make the reader understand what the paper provides 
and whether it could be relevant to his problem.

\item When referencing software packages: The platform(s) they are 
available for.  In which form(at) the results are given, \eg symbolic 
expression, subroutine, table of numbers.

\item The structure the solution possesses.  What kind of functions it 
is comprised of.  What range of validity it has.

\item Whether the entry represents linear combinations, sets of 
integrals, or non-standard propagators.  If so, please also fill the 
dedicated field for the number of master integrals considered in the 
reference.

\item The normalization used in the reference.  For example, this 
normalization will often include factors like $\mu^{4-D}/\Pi$ from the 
integral measure, where it is relevant whether $\Pi$ is defined as 
$\ri\pi^{D/2}$ or $(2\pi)^D$ (or something else).
\end{itemize}
The input fields allow some minor formatting:
\begin{itemize}
\item special characters are entered in HTML, \eg ``\verb=&pi;='' for 
  $\pi$,
\item text in \verb=$...$= is displayed in a font mimicking TeX's math 
  mode,
\item the \verb=b= in \verb=A_b=, \verb=A_{b}=, \verb=A^b=, \verb=A^{b}=
  is sub- (\verb=_=) or superscripted (\verb=^=),
\item URLs and arXiv references are automatically made clickable.
\end{itemize}
If you have machine-readable results, \eg Mathematica notebooks or FORM 
files, upload them even if they duplicate the results in the paper 
cited.

\begin{figure}[ht]
\centerline{\includegraphics[width=.9\hsize]{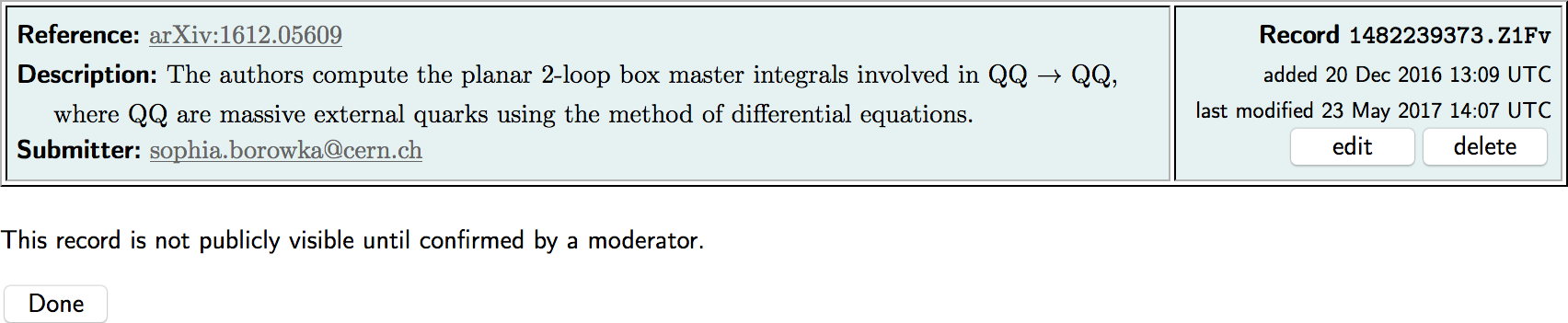}}
\caption{\label{fig:review}The Loopedia Record Review}
\end{figure}

Upon successful submission the submitted record is displayed again for 
review, with controls for editing and deletion (Fig.~\ref{fig:review}).  
Simultaneously an e-mail is sent to the submitter's address (if given) 
with a individualized URL that allows to access the review page again 
for future editing.  While it is possible to submit anonymously we 
strongly encourage submitters to leave their e-mail addresses in case of 
questions or disputes.

The submitted records are not publicly visible until confirmed by a 
moderator.  This is a safety policy we implemented for now to forestall 
arguments, such as submitters deleting each other's records, and may 
change in the future.  The moderators are notified together with the 
submitter and will usually make new records public in a timely manner.  
Moderators may make changes at their discretion, though the present 
editing policy is that substantial changes require contacting the 
submitter, if an e-mail is given.  If you wish to become a moderator 
please contact \texttt{loopedia@mpp.mpg.de}.

The records belonging to a particular CNickel index are kept in one of 
three `visibility' bins, \texttt{private}, \texttt{public}, and 
\texttt{deleted}, where (non-privileged) database lookups only access 
records in \texttt{public}.  A new submission starts life in 
\texttt{private}, moderator confirmation moves it to \texttt{public}, 
deleting it to \texttt{deleted}, and a submitter edit back to 
\texttt{private}.  The moderators' superpowers consist in being able to 
(a) view all bins, (b) move records freely between the bins, and (c) 
remove them completely, too.  All of these actions are logged.

%========================================================================

\subsection{Multiple Upload}

A situation commonly encountered in practice is that several integrals 
are computed in one paper and when adding those one would of course like 
to avoid re-typing the reference, authors, description, etc.  Loopedia 
aids this in two ways.

Firstly by `remembering' information: a record once entered is `carried 
over' to the next graph, \ie the submission form comes with the fields 
already filled out.  Unless one wants to add details specific to the new 
graph, one can just click \button{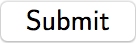}.

Secondly, Loopedia has a special mode for multiple uploads, found in the 
top right corner of the start page: \button{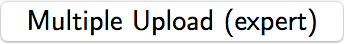}\,.  It requires that 
the CNickels of the graphs are known (though not necessarily 
normalized), or alternately an edge list with the mass identifiers as 
third members is given, as in: \texttt{((e,0,p) (0,1,1) (0,1,1) 
(e,1,p))}.  This is particularly attractive if a program exists to 
generate those.

\begin{figure}[ht]
\centerline{\includegraphics[width=.8\hsize]{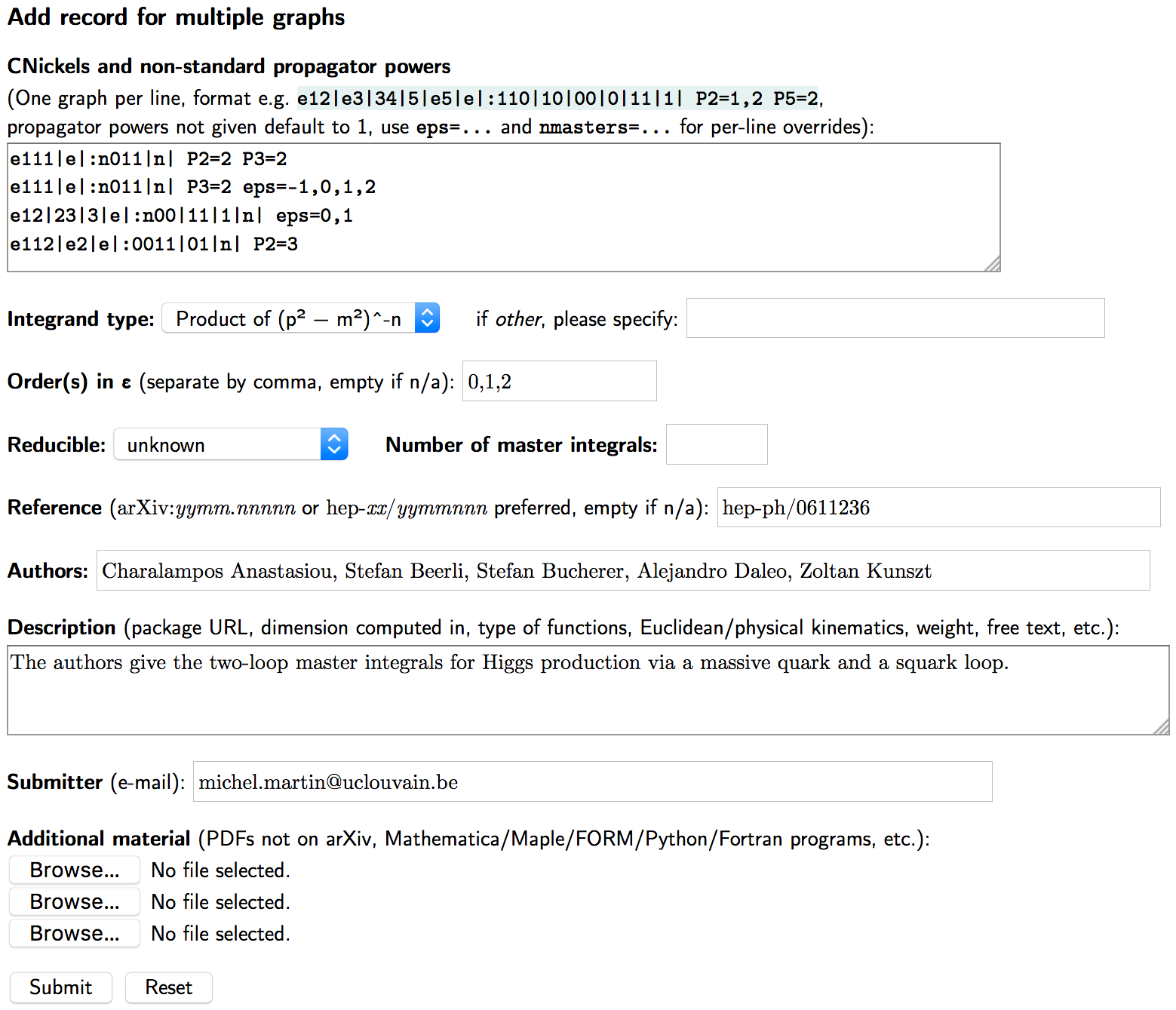}}
\caption{\label{fig:multiadd}The Multiple Upload Form}
\end{figure}

The form for multiple submission is pretty much the same as the standard 
New Record Form except that the graph and any non-standard propagator 
powers are entered in the box at the top (Fig.~\ref{fig:multiadd}).  Per 
line one graph is entered, followed by any non-standard powers in the 
form \texttt{P5=2}.  The $\varepsilon$-order and the number of masters 
can be overridden for that line by respectively adding \eg 
\texttt{eps=-2,-1} or \texttt{nmasters=15}.

\begin{figure}[ht]
\centerline{%
\includegraphics[height=.5\hsize]{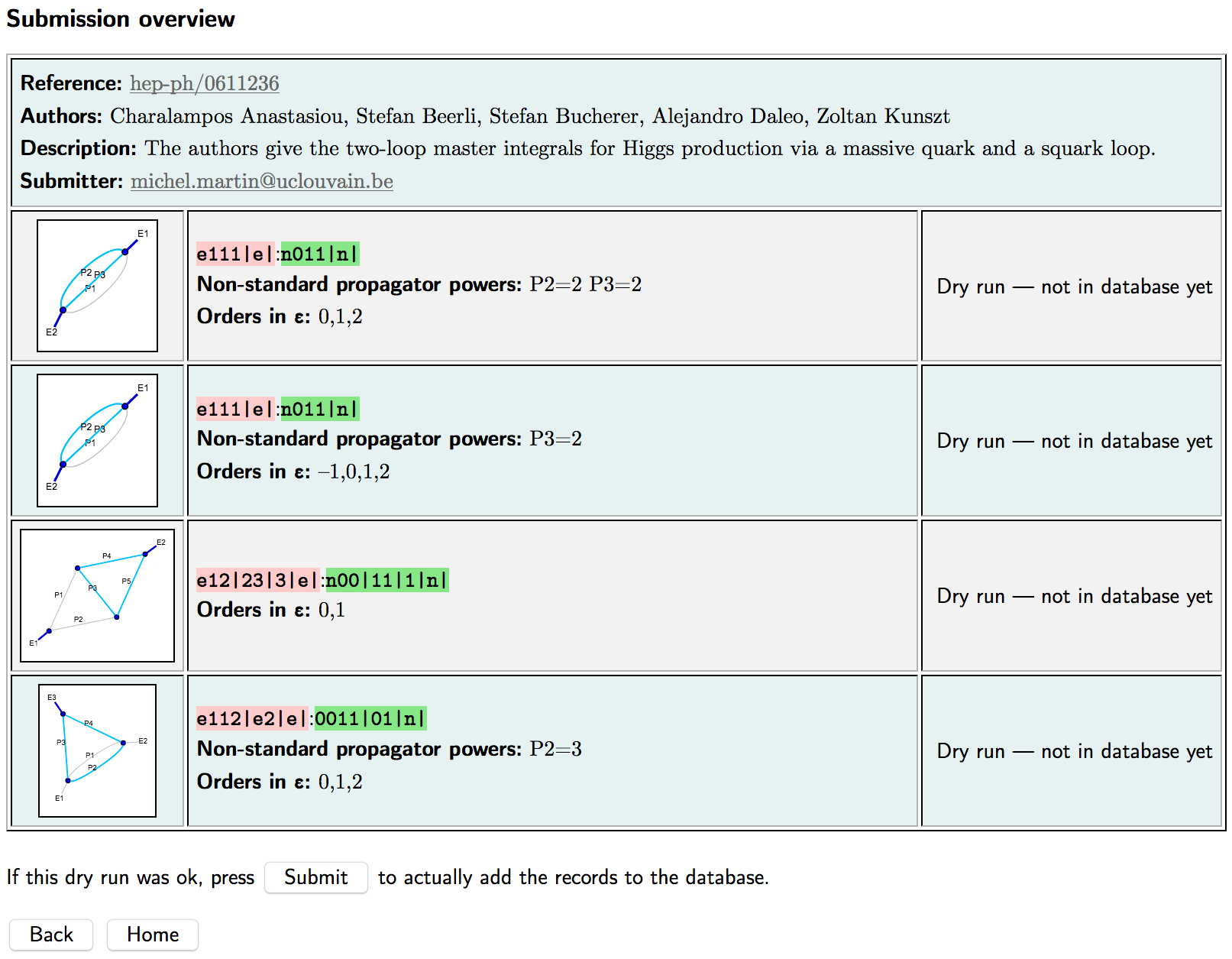}~%
\includegraphics[height=.5\hsize]{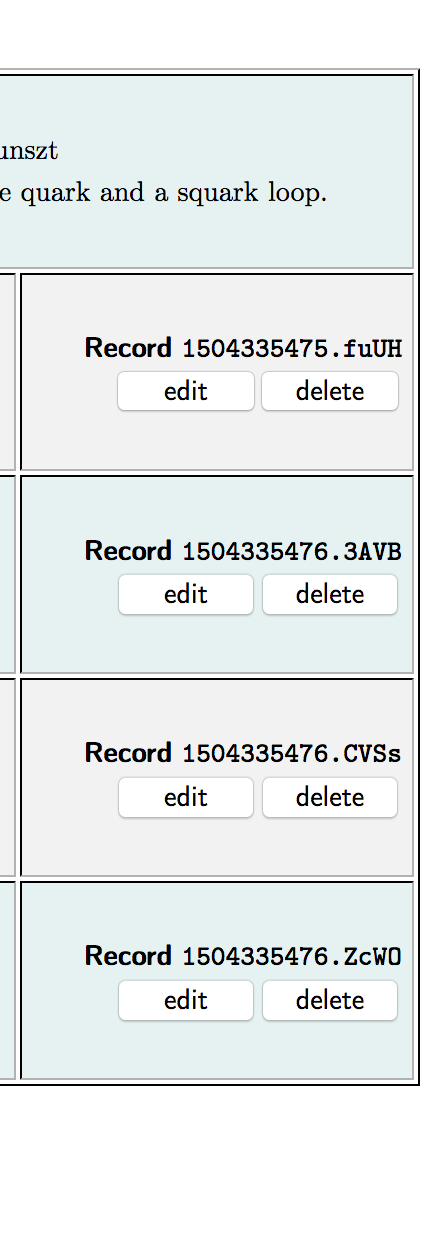}}
\caption{\label{fig:multidry}Dry Run and Final Submit for Multiple Upload}
\end{figure}

The submit button does not immediately add the records to the database 
in this case but performs a `dry run' first (Fig.~\ref{fig:multidry}).

%========================================================================

\section{Internals}

\subsection{Design Decision}

Initially we considered a fairly `standard' design with an SQL 
backengine interfaced through a PHP-based frontend but moved away from 
this design for several reasons, most importantly long-term 
maintainability.  In a science world where software projects often 
languish after the student who wrote them leaves the field we wanted 
our database and frontend to fulfill three requirements:
\begin{itemize}
\item \emph{Maintenance and portability to a different platform should 
  be simple and require only minimal system administration.}

SQL for one needs quite some administrator intervention to set it up 
because it brings along its entire ecosystem.  And at least binary 
portability of the tables is not a given as different 
systems/distributions come with different SQL implementations.

\item \emph{The database entries should be accessible by standard Unix 
  tools.}

This makes inspection and fixing of problems much simpler, and 
furthermore it is less of a burden to the backup since, after each 
modification, only the files that actually changed need to be backuped, 
not the entire SQL table.

\item \emph{The realization should not be based on a software framework 
  which requires frequent updates, in particular security ones.}

For example, one framework we looked at was Drupal, with two major 
versions being maintained and a projected lifetime of 8--10 years for a 
release.  Drupal versions are in general not upward compatible; assuming 
we joined somewhere in the middle of a release, this means having to 
rewrite after about 5 years, not counting updates in relevant 
third-party packages, which typically have no life-cycle management.

\end{itemize}
Finally we settled on a bash script in a CGI environment which uses the 
Unix file system as database, indexed by the \texttt{mlocate} utility, 
and fulfills all three criteria above.

%========================================================================

\subsection{Database Structure}

\begin{figure}[ht]
\centerline{\includegraphics[width=.6\hsize]{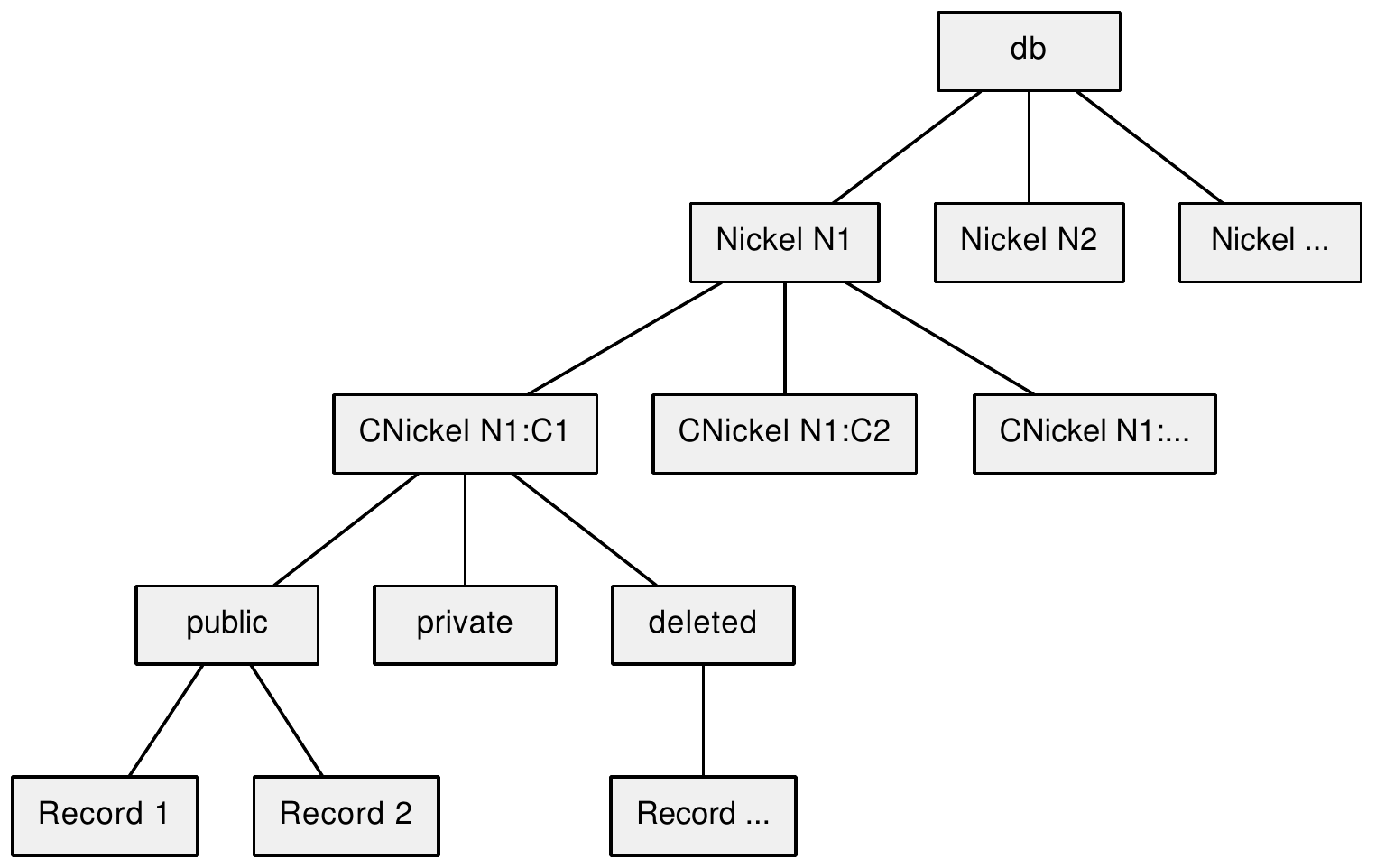}}
\caption{\label{fig:dbstruct}Structure of the Loopedia database.}
\end{figure}

The Loopedia database contains (in a very real filesystem sense) one 
directory for each bare Nickel index, with subdirectories for the 
configuration index (the second part of the CNickel) of the same graph.  
Underneath comes another layer of subdirectories for the visibility 
(\texttt{public}, \texttt{private}, \texttt{deleted}), and below that, 
one directory for each record (Fig.~\ref{fig:dbstruct}).  Doubling the 
filesystem as a database is not all that unusual; Apple, for example, 
chose a similar design for its iTunes music database.

The actual implementation inserts a few more directory layers for 
indexing and performance reasons so that the full database path 
becomes
$$
\texttt{db/} L \texttt{/} \ell \texttt{/} p \texttt{/} \text{Nickel}
\texttt{/} s \texttt{/} \text{Config} \texttt{/} \text{Visibility}
\texttt{/} \text{Record}
$$
where $L$, $\ell$, $p$, $s$ are the number of loops, legs, propagators, 
and scales, respectively.  The database root `\texttt{db}' is included 
in the path to be able to make anchored searches.

Records have identifiers similar to Unix's maildir format, \eg 
\texttt{1482239373.Z1Fv}, which is the Unix time of submission (seconds 
after 1/1/1970) followed by a \texttt{mkdtemp} suffix to avoid race 
conditions.

Indexing of the database is performed with the \texttt{mlocate} Unix 
utility.  The latter indexes the file name only but since that includes 
the extra directory layers just mentioned we can also search for the 
number of loops, for example.  With a little tweak in the source code we 
were able to make \texttt{mlocate} work with relative paths, which means 
its output is exactly of the form above and the index file is 
independent of the database's location in the file tree.

Unix filesystems are actually pretty powerful databases but are neither 
made for nor really expected to scale in extreme cases.  A common 
bottleneck is excessively many files in one directory -- a case which in 
Loopedia is efficiently mitigated by the extra directory layers 
introduced above.  Loopedia is currently running on Linux's Ext4 
journaling filesystem.

%========================================================================

\subsection{Graph-theoretical Operations}
\label{sect:graphth}

Almost all graph-theoretical operations in Loopedia are taken care of by 
the Python library GraphState \cite{graphstate}.  GraphState allows to 
supplement a graph with configurations, \ie add extra integers to either 
the nodes or the edges.  The colored Nickel index used by Loopedia 
implements a single edge coloring which encodes masses and external 
$q^2$.

GraphState's default output format for a configuration is decimal, with 
underscores delimiting the integers, for example 
\verb=0_0_0_0|0_0_0|38|=.  We wanted the coloring to have the same 
length as the bare Nickel index, however, since that makes visual 
pairing of the bare Nickel index (left column in Graph Browser) and its 
colorings (right column) particularly straightforward, and also 
identifiers like `\texttt{38}' (which might stand for `any mass scale') 
are not very self-explanatory.  Therefore we generalized GraphState's 
formatting routines slightly, to allow the CNickel notation described 
earlier.

GraphState automatically brings each graph into a normal form but does 
not touch the coloring as it has no notion of what it represents.  We 
therefore added a routine to canonicalize the entire CNickel so that no 
spurious duplicates would swamp the database.  Operations include: a 
tadpole's single leg is cut off since it carries no information for the 
integral, single instances of definite scales (\texttt{1}, \texttt{2}, 
\dots) are converted to `any non-zero scale' (\texttt{n}), remaining 
definite scales are renumbered, for two-point functions momentum 
conservation is enforced (\ie equal identifiers for incoming and 
outgoing legs).

Lastly, we added a Python function to process user input and bring it 
into the form of an edge list acceptable to GraphState.  Mostly we strip 
function heads (\ie replace $f(i)$ by $i$) and if the remaining list has 
no substructure, partition it in pairs of two.  This means that the 
input for the same diagram can be as diverse as
\begin{itemize}
\item \texttt{1 3 2 4 3 4 3 4}
\item \texttt{[(1,3) (2,4) (3,4) (3,4)]}
\item \begin{verbatim}
Topology[2][
  Propagator[Incoming][Vertex[1][1], Vertex[3][3]], 
  Propagator[Outgoing][Vertex[1][2], Vertex[3][4]], 
  Propagator[Loop[1]][Vertex[3][3], Vertex[3][4]], 
  Propagator[Loop[1]][Vertex[3][3], Vertex[3][4]] ]
\end{verbatim}
\end{itemize}
A QGRAF output style suitable for Loopedia would be
\begin{verbatim}
<prologue>
<diagram>
<propagator_loop>(<vertex_index>,<dual-vertex_index>) <end>
<epilogue>
<exit>
\end{verbatim}

%========================================================================

\subsection{Drawing Graphs}

The most appealing representation of a graph for humans is still its 
picture, and indeed both the Graph Browser and the Single-Graph Display 
take advantage of this.  Automatically drawing an arbitrary graph (with 
reasonable output) is a difficult business, however, which we gladly 
leave to the `neato' component of the Graphviz package \cite{graphviz}. 
The shapes may not always be the ones traditionally associated with 
Feynman diagrams but the important point here is that the propagator 
routing and labelling can be understood at a glance with as few 
ambiguities as possible.

Behind the scenes a Python function translates the graph into Graphviz's 
DOT language, from which neato produces an SVG image.  SVG is an 
XML-based vector format rendered by all modern browsers and scales well 
in the icon-size plaquettes of the Graph Browser.  All images are laid 
out with the same absolute font sizes and length scales but may appear 
with different weights relative to each other in the Graph Browser due 
to being scaled to fit.

The graph corresponding to a colored Nickel index is really colored.  In 
particular the massless case, drawn in grey with reduced line weight, 
can instantly be recognized even in the small image sizes of the Graph 
Browser.

%========================================================================

\subsection{HTTP and CGI handling}

The `heart chamber' of Loopedia is a bash script named 
\texttt{index.cgi}.  It interacts with the Web server through the Common 
Gateway Interface (CGI).  Also CGI is known to suffer from common 
exploits through unguarded handling of the CGI query strings but we 
have two tough little C programs for processing.

For the default type of HTML form 
(\texttt{application/x-www-form-urlencoded}) we use our 
\texttt{unescape} utility, which parses the CGI input into variable 
assignments suitable for bash's \texttt{eval} command.  Needless to say, 
the left-hand sides are suitably sanitized and the right-hand sides 
properly quoted to prohibit execution of remote code.  

HTML forms emanating from \texttt{index.cgi} submit only with the 
\texttt{POST} method, mainly so as not to mess up the user's URL field, 
but in the e-mail to submitter and moderators a \texttt{GET}-type URL is 
sent for direct referral to a particular record, and of course 
\texttt{unescape} can deal with any combination of \texttt{POST} (stdin) 
and \texttt{GET} (environment) input.

For the record submission we must use the \texttt{multipart/form-data} 
type since it may include upload of (binary) files.  This kind of input 
is handled by \texttt{formdecode}; it splits the input stream along the 
delimiters and stores the results in files, in a temporary directory.  
Obviously we have to sanitize these filenames, too, but want to keep the 
original form as much as security allows, for there is usually some 
meaning at least in the extension (\texttt{.pdf}, \texttt{.f}, 
\texttt{.m}) if not in the entire name.

Except in the e-mails, \texttt{index.cgi} at no point references an 
absolute URL -- in fact, the HTML forms are completely self-referential 
as there is only one CGI script around.  Not only does this make testing 
easy (the script runs without change on `localhost' on a developer's 
laptop, for example) but we can also manage privileged access simply by
symlinking the Loopedia root directory to a different directory for 
which we require authentication through the Apache configuration.  
Elevated rights are granted if the server presents a valid user.

Submitters gain restricted privileges to modify their own submission 
through a URL including a token.  The random 32-character alphanumeric 
token is generated at submission and e-mailed to the submitter.  It both 
identifies the record it belongs to uniquely and grants permission to 
edit or delete that record.

Loopedia's outward appearance is governed by the \texttt{loopedia.css} 
style sheet and degrades gracefully if the CSS takes too long to load or 
is blocked.  CSS and the generated HTML code validate cleanly (no 
errors, no warnings) on $\{\text{Firefox}, \text{Chrome}, \text{Safari}, 
\text{Opera}\}\otimes \{\text{Linux}, \text{MacOS}\}$.

%========================================================================

\section{Summary}

In this paper we presented a new database for bibliographic and other 
information on loop integrals.  Loopedia collects the available 
information and makes it searchable by graph.  We hope that Loopedia 
will in time be able to answer the query ``Find all papers pertaining to 
graph $X$.''

The database is hosted at the Max Planck Institute for Physics in Munich 
with broadband internet access and a daily backup at 
\texttt{loopedia.org}.

It is now up to the loop-integral calculators to make this effort a 
success -- and \textit{en passant} attract more citations for their 
work.

%========================================================================

\section*{Acknowledgements}

We thank Claude Duhr and Valentin Hirschi for important discussions. 
C.~Bogner thanks Deutsche Forschungsgemeinschaft for support under the 
project BO 4500/1--1.  S.~Borowka gratefully acknowledges financial 
support by the ERC Advanced Grant MC@NNLO (340983) and ERC Starting 
Grant ``MathAm'' (39568) during different stages of this project.  
M.~Michel is supported by the ``Fonds Sp\'ecial de Recherche'' of the 
Universit\'e catholique de Louvain and grateful to the TH department at 
CERN for hospitality during various stage of this project.  This 
research was supported in part by the Research Executive Agency (REA) of 
the European Union under the Grant Agreement PITN-GA2012316704 
(HiggsTools).  We gratefully acknowledge support by the Munich Institute 
for Astro- and Particle Physics (MIAPP) of the DFG cluster of excellence 
``Origin and Structure of the Universe.''

%========================================================================

\end{document}